\documentclass[11pt,twoside]{article}
\pdfoutput=1
\usepackage{asp2014}

\aspSuppressVolSlug
\resetcounters

\begin{document}

\newcommand{\arxiv}[1]{\href{http://arxiv.org/abs/#1}{\texttt{arXiv:#1}}}
\newcommand{\procspie}{Proc.\ SPIE}

\title{Porting the LSST Data Management Pipeline Software to Python 3}
\author{Tim~Jenness
\affil{Large Synoptic Survey Telescope, Tucson, AZ, USA; \email{tjenness@lsst.org}}}

% This section is for ADS Processing.  There must be one line per author.
\paperauthor{Tim~Jenness}{tjenness@lsst.org}{0000-0001-5982-167X}{LSST}{Data Management}{Tucson}{AZ}{85719}{USA}

\begin{abstract}
The LSST data management science pipelines software consists of more than 100,000 lines of Python 2 code.
LSST operations will begin after support for Python 2 has been dropped by the Python community in 2020, and we must therefore plan to migrate the codebase to Python 3.
During the transition period we must also support our community of active Python 2 users and this complicates the porting significantly.
We have decided to use the Python \texttt{future} package as the basis for our port to enable support for Python 2 and Python 3 simultaneously, whilst developing with a mindset more suited to Python 3.
In this paper we report on the current status of the port and the difficulties that have been encountered.
\end{abstract}

\section{Background}

The Large Synoptic Survey Telescope \citep[LSST;][]{2008arXiv0805.2366I} will take about 15\,TB of image data per night and after ten years of operations will have 15\,petabytes of catalog data for the final data release and 0.5\,exabytes of image data\footnote{\url{http://lsst.org/scientists/keynumbers}}.
As part of the LSST Data Management System \citep{O3-1_adassxxv}, we are writing a suite of software packages to enable these data products to be created with sufficient quality and performance to meet the established science goals \citep{2009arXiv0912.0201L}.

Development of the LSST pipeline software began in 2004 \citep{2004AAS...20510811A}, when Python was at version 2.3, and has continued through the research and development phase \citep{2010SPIE.7740E..15A} into the construction phase \citep{P056_adassxxv}.
During these 12 years the code base has migrated through different versions of Python 2 but had not been made compatible with Python 3 until this year.
In 2004 the expected first light for LSST was 2012 \citep{2004SPIE.5489..705C}, and the release of the first ``usable'' version of Python 3 (3.2) was not made until 2011, so during the early years of project development there was no need to worry about future Python developments.
Now the astronomy computing environment is beginning to change and Python 3 is becoming more familiar to the community\footnote{\url{http://astrofrog.github.io/blog/2015/05/09/2015-survey-results/}}.
There were some key motivators for LSST to support Python 3: (a) The official statement from the Python developers is that Python 2.7 support will be dropped in 2020.
This is before the official start of the LSST survey and we do not want to commission software where the key executable under-pinning the entire system will soon lose support.
(b) LSST has external users of our software that provide early beta testing services and we do not want to actively impede those users from migrating to a more modern Python.
(c) Astropy \citep{2013A&A...558A..33A} recently declared that it would do bugfix-only maintenance releases supporting Python 2 after 2017, and that subsequent major releases would be Python 3 only \citep{APE10}, joining IPython, pandas, sunpy and matplotlib. This declaration will motivate the community, and furthermore, LSST DM recently decided to integrate Astropy into our pipelines code \citep{doi:10.1117/12.2231313}.

\section{Supporting Python 3}

A key requirement for this initial port was that our pipeline code that is used by external users must support both Python 2.7 and Python 3.
The Python community has developed a number of schemes for handling this and we looked at both \texttt{six} (used by Astropy) and \texttt{future}.
We decided on \texttt{future} because the resulting code looks almost exactly like Python 3 code, in many cases the code can run on Python 3 without \texttt{future} being installed, and the \texttt{futurize} command provides an easy migration path.

Before modernizing the code, we took the opportunity to tidy it using the \texttt{autopep8} tool to fix simple whitespace inconsistencies.
This includes replacing tabs with spaces as Python 3 no longer allows a mix of tab and space indents.
The eventual aim is to run the \texttt{flake8} tool on all code submissions to allow code reviewers to focus on architectural and algorithmic issues rather than being distracted by coding standard violations.

The next step was to run the \texttt{futurize} stage 1 command in order to modernize the codebase to use Python 2.7 features.
This step adds no additional dependencies but is an important modernization required before supporting Python 3.
In our code the main modernizations were enabling print as a function and absolute import from \texttt{\_\_future\_\_}, using the \texttt{in} operator rather than \texttt{has\_key()} and also modernizing exception handling to use the ``\texttt{except Exception as e}'' syntax rather than the older ``\texttt{except Exception, e}'' form.

The final automated phase is to run the stage 2 \texttt{futurize} command.
This command is used to make an initial pass on Python 3 compatibility by scanning each file to determine which compatibility imports are need.
The \texttt{2to3} tool is used internally, but with the addition of new shim imports that are no-ops on Python 3 but which change the behavior of builtins on Python 2 to match those in Python 3.
Additionally, we had a number of places that used the \texttt{map(filter(lambda...))} construct and these are automatically replaced with list comprehensions.

\section{The Experience}

\paragraph{Lists versus Iterators}

The \texttt{futurize} command is very defensive.
If it sees a construct that resulted in a \texttt{list} in Python 2 but which now returns a view or iterator, the command will wrap it with a \texttt{list}.
In some cases this is correct, but in many cases it is suboptimal, therefore some amount of inspection is required to decide on each case.
A common example is where a loop is written as ``\texttt{for i in a.keys():}''.
This is converted to ``\texttt{for i in list(a.keys())}:'' rather than the more idiomatic ``\texttt{for i in a:}''.

\paragraph{Bytes versus Characters}

The main difficulty in supporting Python 2 and 3 is that Python 2 has a very relaxed attitude to bytes, characters and Unicode.
Python 3 has very well-defined semantics distinguishing bytes from characters and any code that deals with the output from external commands (\texttt{subprocess.check\_output} for example) or that needs to distinguish binary files from text files (in our case \texttt{pickle} files) must properly decode the bytes to characters.
In some cases the function is really working with bytes and does not want characters at all.

One complication we had revolved around our SWIG \citep{beazley2003automated} interface.
By default in Python 2 SWIG treats bytes as \texttt{std::string} and does not map \texttt{unicode} strings.
We enabled the setting to allow \texttt{unicode} to also map to \texttt{std::string} but in some cases our C++ APIs were expecting bytes and had to have explicit SWIG interfaces created to prevent them being treated as valid Unicode byte representations.
Sometimes this assumption triggered a segmentation fault.

\paragraph{\texttt{str}}

A Python 3 \texttt{str} is a Unicode string, similar to a Python 2 \texttt{unicode}.
The \texttt{future} package provides a \texttt{str} that can be used on Python 2 that emulates that found on Python 3, and for our initial migrations we accepted these changes from the \texttt{futurize} command.
As we convert more code, we are realizing that since we are not using any Unicode features, switching to a Python 3-compatible \texttt{str} is not important.
Furthermore, it is actively breaking code.
For example, unless \texttt{unicode\_literals} is enabled, literal strings in Python 2 are standard Python 2 \texttt{str} objects but the strings created with \texttt{str} are not.
If some of the code is using native \texttt{str} but the rest is using \texttt{future} \texttt{str} any use of \texttt{isinstance(var, str)} can return \texttt{False} even if the supplied variable looks like a string to the developer.
To overcome some of these confusions on Python 2 \texttt{basestring} is required in \texttt{isinstance} calls and this must be imported from \texttt{past.builtins} to allow the code to work on Python 3.
We may have to reassess our usage of the \texttt{future} \texttt{str}.

\paragraph{\texttt{long}}

Much of our code was explicitly using the \texttt{long} integer type, both as a constructor (\texttt{long()}) and for literals (e.g.\ \texttt{0L}).
Python 3 does not support the \texttt{L} syntax and these must be removed and be replaced with \texttt{long()}.
In some cases we decided to use the \texttt{future} package \texttt{past.builtins} to import a \texttt{long()} constructor in Python 3 (an alias of the normal \texttt{int()}) whilst retaining the standard Python 2 semantics.
In reality, this was not required since our code was using \texttt{long()} to indicate a 64-bit integer, and on 64-bit systems, Python 2 uses 64-bit integers (our code base used to run on 32-bit but that is no longer supported).
It took us a while to realize this, and we have started removing the explicit use of \texttt{long()}.
This has led to us also clarifying some of our Python/C++ interfaces to remove the assumption that a Python \texttt{int} type maps to a 32-bit C++ integer and a Python \texttt{long} maps to a 64-bit C++ integer.

\paragraph{Version-dependent Code}

The success of the \texttt{future} abstraction can be seen in how few places there are in the code that require specific Python version tests.
Currently, there is one place where the \texttt{encoding} argument is needed in \texttt{pickle.load} to load a Python 2 \texttt{pickle} file.
There is one test that is disabled on Python 2 because Python 2 can not indicate to the C++ interface that there is any difference between bytes and characters.
There is one place that needs to know whether a function is associated with a builtin (\texttt{\_\_builtins\_\_} on 2 and \texttt{builtins} on 3).
The most complex problem we encountered was handling classes that use multiple inheritance of two classes which themselves define their own metaclasses (in our case these were \texttt{collections.UserDict} and \texttt{yaml.YAMLObject}.
This is not allowed in Python 3 without defining a new metaclass that derives from both the relevant types.
The syntax differences meant that the class definitions had to occur in versioned code branches.
The imports from \texttt{past.builtins} will be removed when support for Python 2 is dropped (replacing \texttt{basestring} with \texttt{str} and \texttt{long} with \texttt{int}).

\section{Current Status}

As of October 2016, approximately 45 LSST Science Pipelines packages have been converted to support both Python 2 and Python 3, with about 10 packages still remaining.
We have Jenkins continuous integration jobs running daily on multiple operating systems and multiple Python versions to ensure that Python 2-specific code does not slip back into the codebase.
More detailed instructions on our process can be found in \citet{SQR-014}.

We hope to complete this work before the end of the year.
Other Data Management code, including Qserv \citep{Wang:2011:QDS:2063348.2063364} and the data access libraries, will be migrated over the coming winter.
We may decide to drop Python 2 support for these as it is mostly server code and the migration to Python 3 is made significantly easier if support for 2 is not required.

\acknowledgements I thank Russell Owen and Fred Moolekamp for comments on the draft.
The LSST software stack is the result of the efforts of the many people who are part of the Data Management Team at LSST, as well as outside contributors.
This material is based upon work supported in part by the National Science Foundation through Cooperative Agreement 1258333 managed by the Association of Universities for Research in Astronomy (AURA), and the Department of Energy under Contract No. DE-AC02-76SF00515 with the SLAC National Accelerator Laboratory. Additional LSST funding comes from private donations, grants to universities, and in-kind support from LSSTC Institutional Members.

\end{document}